\documentclass[aps,pdftex,twocolumn,groupedaddress,superscriptaddress,showpacs,amsmath,amssymb,floatfix,pra,showkeys,raggedbottom]{revtex4-1}

\pdfoutput=1 
\usepackage{dcolumn}
\usepackage{graphicx} 
\usepackage{bm}
\usepackage{wasysym}
\usepackage{mathrsfs}
\usepackage[normalem]{ulem}
\usepackage{epstopdf}

\usepackage{natbib,hyperref}

\usepackage[usenames,dvipsnames]{xcolor}

\usepackage{mdframed}
\usepackage{srcltx}

\newcommand{\subfigimg}[3][,]{%
  \setbox1=\hbox{\includegraphics[#1]{#3}}
  \leavevmode\rlap{\usebox1}
  \rlap{\hspace*{3pt}\raisebox{\dimexpr\ht1-\baselineskip}{#2}}
  \phantom{\usebox1}
}

\newcommand{\subfigimgtwo}[3][,]{%
  \setbox1=\hbox{\includegraphics[#1]{#3}}
  \leavevmode\rlap{\usebox1}
  \rlap{\hspace*{0pt}\raisebox{\dimexpr\ht1-\baselineskip}{#2}}
  \phantom{\usebox1}
}

\begin{document}

\title{Two dimensional grating magneto-optical trap}

\author{Eric Imhof}
\email{eric.imhof@sdl.usu.edu}
\affiliation{Air Force Institute of Technology, Wright-Patterson AFB, Ohio, 45433, USA }
\affiliation{Space Dynamics Laboratory, Albuquerque, New Mexico, 87106, USA}
\author{Benjamin K. Stuhl}
\affiliation{Space Dynamics Laboratory, Albuquerque, New Mexico, 87106, USA}

\author{Brian Kasch}
\author{Bethany Kroese}
\author{Spencer E. Olson}
\author{Matthew B. Squires}
\affiliation{Air Force Research Laboratory, Kirtland AFB, New Mexico 87117, USA}

\date{\today}

\begin{abstract}

We demonstrate a two dimensional grating magneto-optical trap (2D~GMOT) with a single input cooling laser beam and a planar diffraction grating using $^{87}$Rb. This configuration increases experimental access when compared with a traditional 2D~MOT.  As described in the paper, the output flux is several hundred million rubidium atoms/s at a mean velocity of 16.5(9) m/s and a velocity distribution of 4(3) m/s standard deviation.  We use the atomic beam from the 2D~GMOT to demonstrate loading of a three dimensional grating MOT (3D~GMOT) with $2.46(7)\times 10^8$ atoms.  Methods to improve output flux are discussed.  

\end{abstract}


\maketitle

\section{Introduction}

Matter wave interferometry has demonstrated orders of magnitude improvement over a wide range of precision measurements \cite{Barrett2011, Cahn1997, Cronin2009, Adams1993, Durfee2006, Dickerson2013, Debs2013, Kovachy2015}.  These successes have spurred interest in transitioning cold atom devices from the lab to more demanding environments \cite{Hogan2011, Imhof2016, Muntinga2013, Rushton2014, Williams2016, Barrett2016, Geiger2011, Hardman2016, Keil2016}.  Recently, a three dimensional grating magneto-optical trap (3D~GMOT) was demonstrated that satisfies many needs of a deployable system \cite{Nshii, Vangeleyn2010, Esteve2013}.  Particularly, the GMOT increases optical access while reducing system size, weight, power, and cost compared to conventional techniques.  

\begin{figure}
  \begin{tabular}{@{}p{0.42\linewidth}@{\quad}p{0.55\linewidth}@{}}
    \subfigimg[width=\linewidth]{(a)}{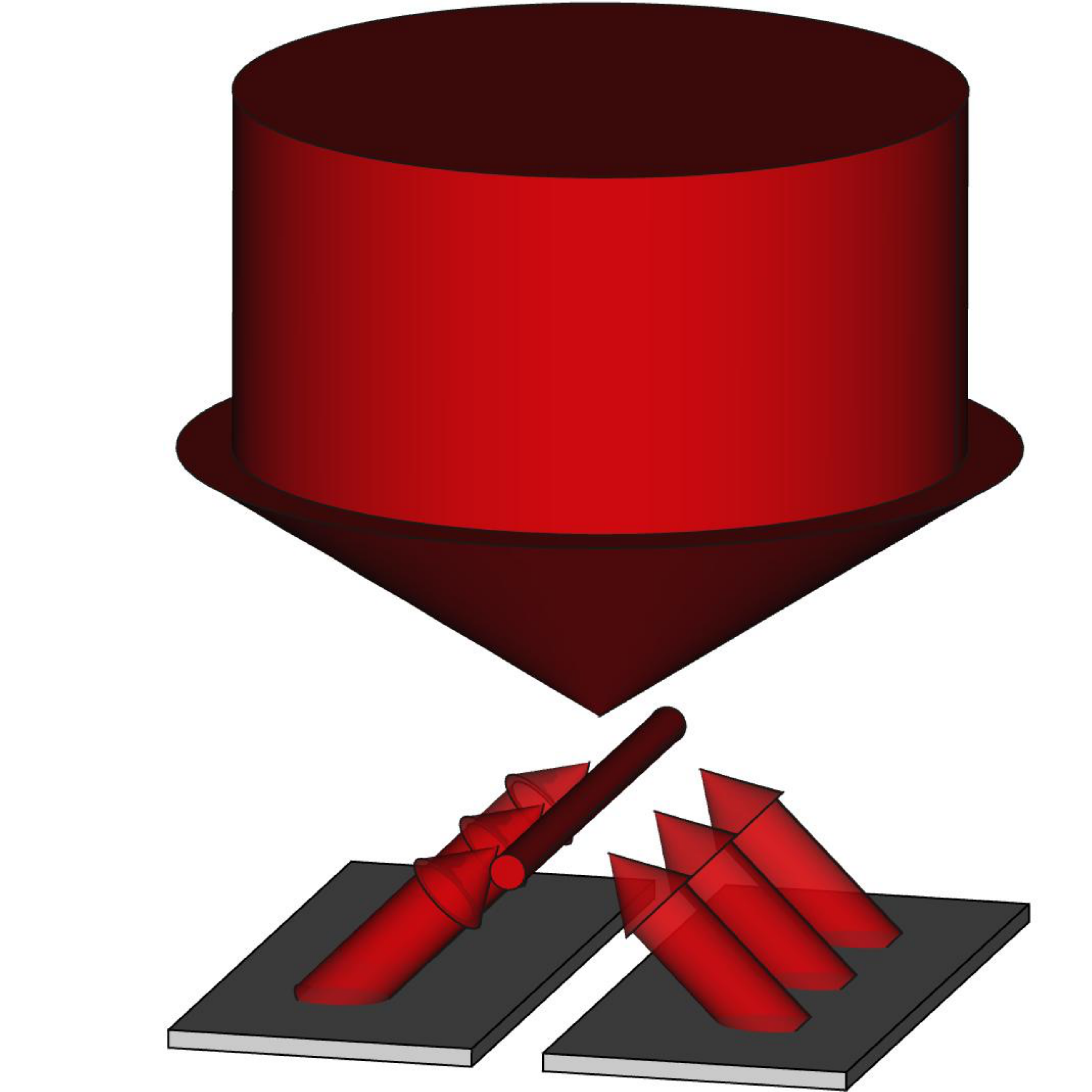} &
    \subfigimg[width=\linewidth]{(b)}{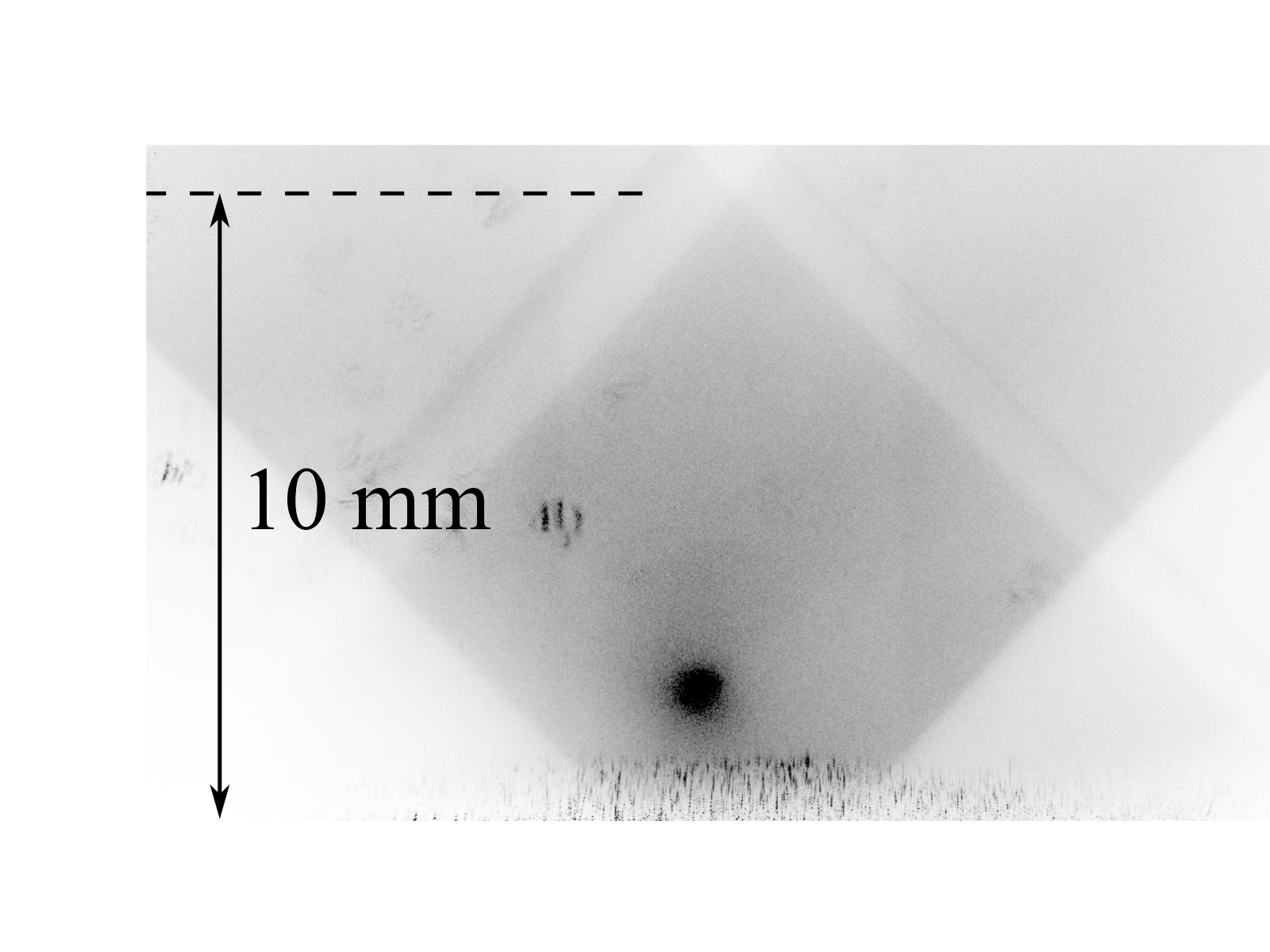} \\
    \subfigimg[width=\linewidth]{(c)}{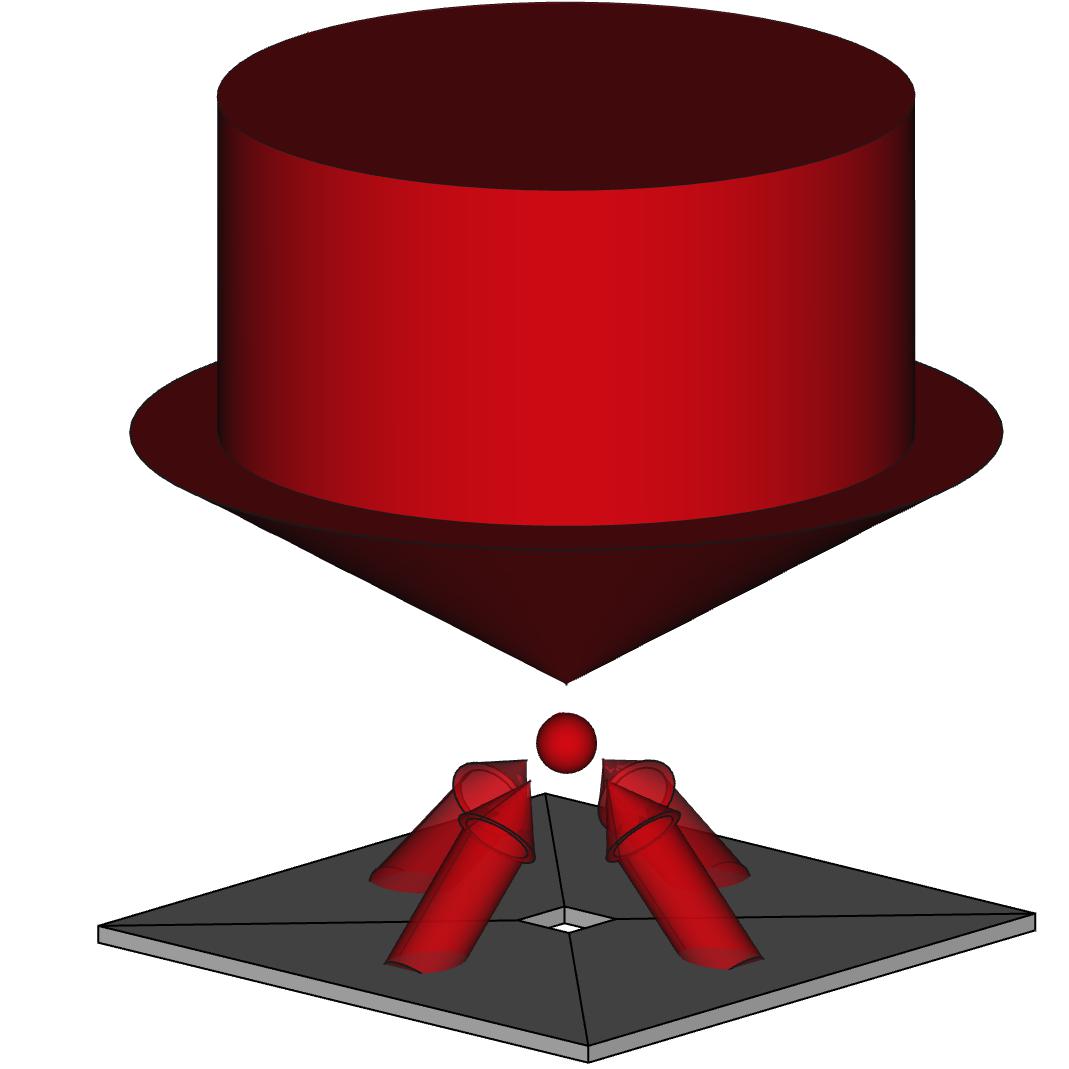} &
    \subfigimg[width=\linewidth]{(d)}{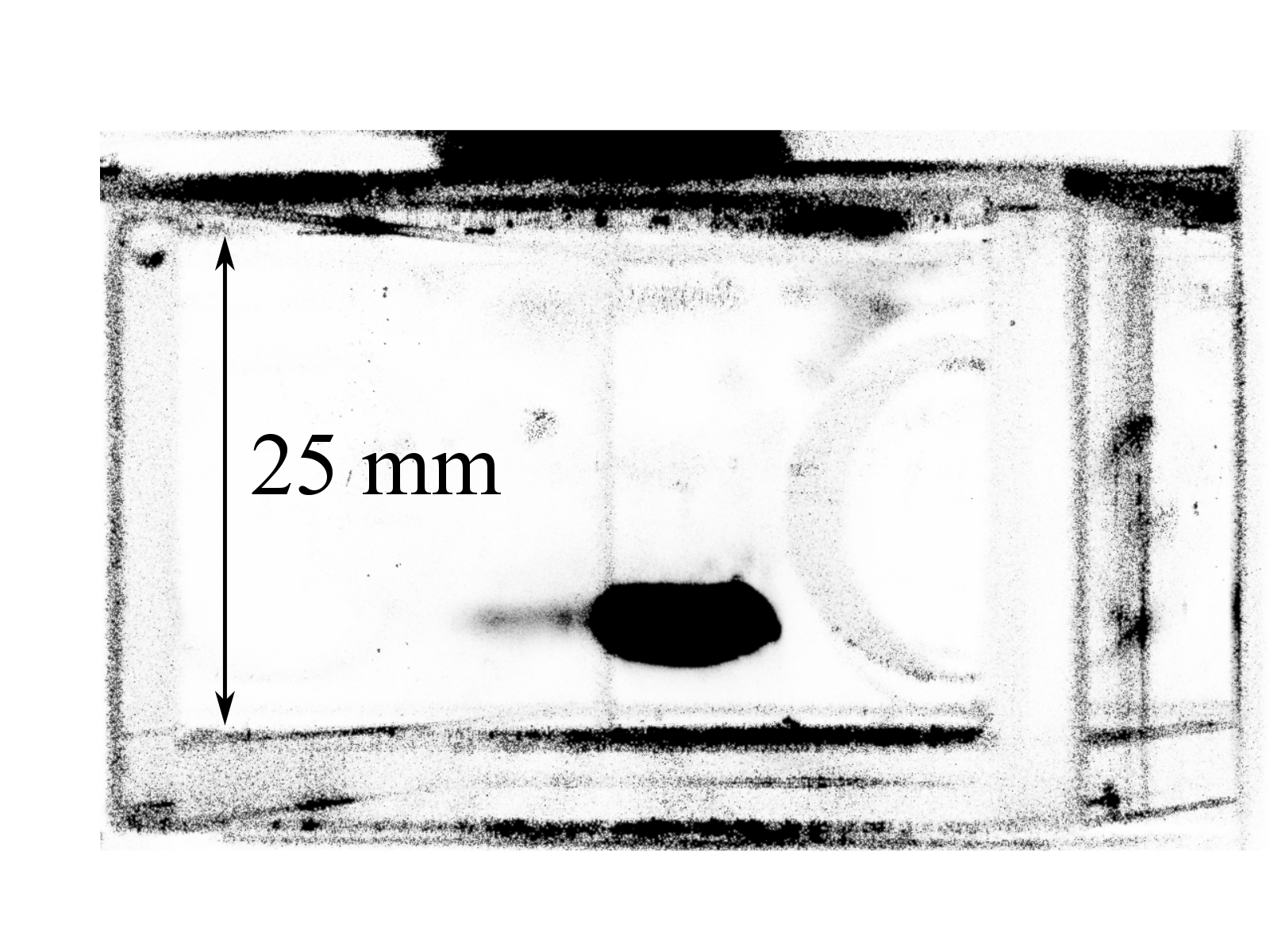} 
  \end{tabular}
  \caption{(Color online) (a) A laser beam impinges on a series of diffraction gratings to form a 2D~GMOT. (b) Inverted greyscale fluorescence of the 2D~GMOT viewed along its axis.  (c) A schematic of a 3D~GMOT and (d) its corresponding inverted fluorescence image.  }
  \label{fig: GMOTSimple}
\end{figure}

A similar principle can be used to form a two dimensional GMOT (2D~GMOT), resulting in a cold atomic beam.  As shown in Fig.~\ref{fig: GMOTSimple}(a)-(b), a 2D~GMOT is formed when a single red-detuned laser beam is normally incident on a pair of planar diffraction gratings.  The diffracted beams intersect with the incident light to provide cooling along two axes.  Assuming proper conditions of polarization and magnetic field, atoms are captured within the region of beam overlap.  

The 2D~GMOT is used to load a 3D~GMOT in a different chamber, shown in Fig.~\ref{fig: GMOTSimple}(c)-(d).  The 2D~GMOT enables faster loading rates and higher atom number in the 3D~GMOT by separating the source vapor from the experimental region. The resulting 3D~GMOT shows comparable atom number scaling to standard six-beam MOT's \cite{Nshii} and is able to achieve sub-Doppler cooling \cite{Lee2013}.  

The rest of the paper will be organized as follows:  the theory considerations for adapting from the 3D to the 2D case will be detailed.  The design and characteristics of a 2D~GMOT with Doppler cooling along the atom beam axis (the 2D$^+$ configuration \cite{Walraven1998}) are then presented.  Finally, the loading rates, lifetime, and atom number of the combined 2D$^+$ to 3D~GMOT system are reported.

\section{Theory and Design} 
Unlike conventional MOT configurations, the GMOT laser beams are not aligned with the magnetic field axes.  Accordingly, specific conditions for intensity and polarization must be considered when selecting gratings.  These conditions differ between the 2D and 3D case.  

Each atom in a MOT scatters light from multiple off-resonant laser beams with wavevectors $\textbf{k}_j$ and polarization vectors $\hat{\boldsymbol{\epsilon}}_j$.  Assuming the atom absorbs from $F = 0\rightarrow F'=1$, a circularly polarized beam drives transitions to the $m_F = -1,0,+1$ excited states with relative strengths $\alpha_{m_F}(\varphi,\hat{\boldsymbol{\epsilon}}_j)$ that depend on the beam's polarization and angle with respect to the local magnetic field $\varphi$.  For a beam whose polarization is labeled by $s = +1$ for right circular or $s = -1$ for left, these strengths are  $\alpha_{\pm 1} = (1\mp s\cos{\varphi})^2/4$ and $\alpha_0 = (\mbox{sin}^2{\varphi})/2$.  The average force from a single beam $j$, of intensity $I_j$, on an atom with velocity $\textbf{v}$ in a magnetic field $\textbf{B}$ is

\begin{equation}
\textbf{F}_j = \hbar \textbf{k}_j\frac{\Gamma}{2} \frac{I_j}{I_{\mathrm{sat}}} \sum_{m_F=-1,0,1} \frac{\alpha_{m_F}(\varphi, \hat{\boldsymbol{\epsilon}}_j)}{1 + \frac{\sum_j I_j}{I_{\mathrm{sat}}} + \frac{4(\Delta - \textbf{k}_j\cdot\textbf{v}-\mu_F m_F B/\hbar)^2}{\Gamma^2}},
\label{eq: fullforcenoapprox}
\end{equation}
where $\Gamma$ is the natural linewidth and $\Delta = \omega_L - \omega_0$, the detuning of the laser frequency from the transition. $I_{\mathrm{sat}}$ is the saturation intensity and $\mu_F = g_F\mu_B$.  In the limit of small Doppler and Zeeman shifts, the force becomes

\begin{equation}
\textbf{F}_j \approx \hbar \textbf{k}_j \frac{\Gamma}{2} \frac{I_j}{I_{\mathrm{sat}}} \left[ K + C\left( \textbf{k}_j \cdot \textbf{v} - \frac{\mu_F s}{\hbar} \frac{\textbf{k}_j \cdot \textbf{B}}{|\textbf{k}_j|} \right) \right],
\label{eq: force2}
\end{equation}
where $K = (1+ \sum_jI_j/I_{\mathrm{sat}} + 4\Delta^2/\Gamma^2)^{-1}$, $C = 8 \Delta K^2/\Gamma^2$\cite{VangeleynThesis}.  

Contrary to common MOT geometries, the optimal light field for a GMOT does not have pure circular polarization because $|\hat{\textbf{k}}_j\cdot\hat{\textbf{B}}| \neq 1$ for the diffracted beams.  In addition, intensity balance states $\sum_j I_j\textbf{k}_j = 0$, requiring beam intensity to change with the diffraction angle.  As a result, the 2D and 3D~GMOT configurations have different constraints, as shown in the following.  

\begin{figure}[htbp]
\includegraphics[width=\linewidth]{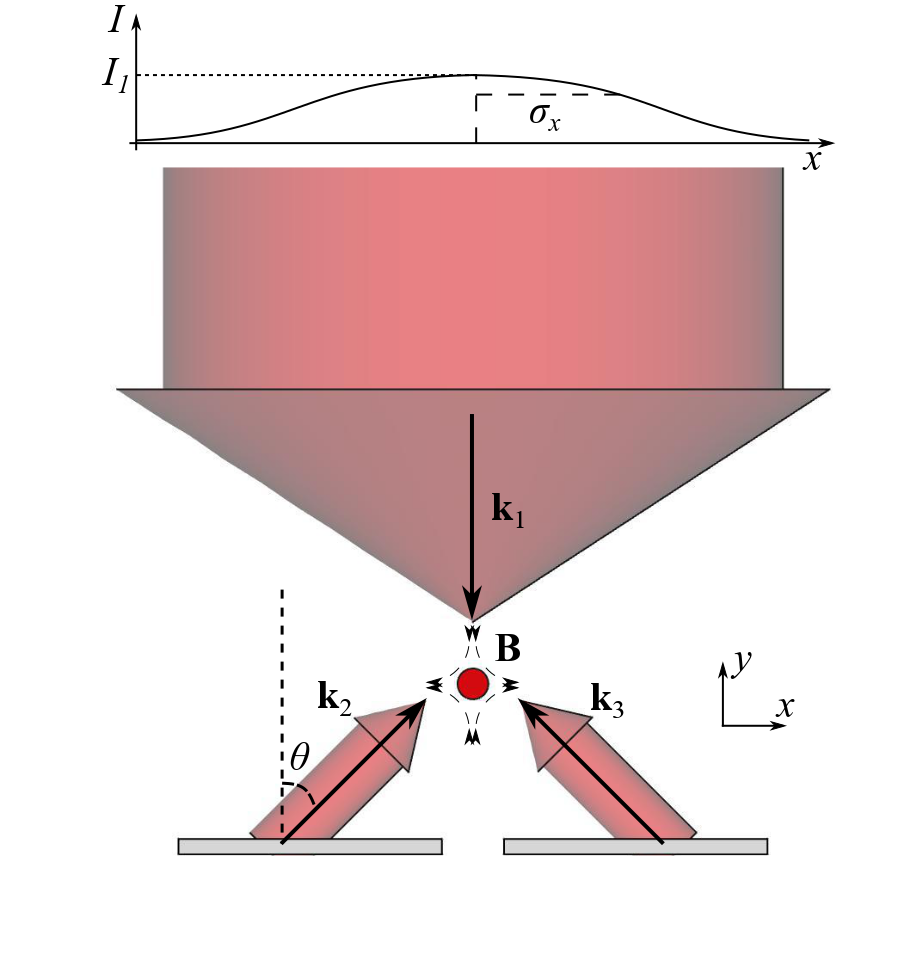}
\caption{(Color online) A single broad input laser beam has direction labeled by its wavevector $\textbf{k}_1$.  The input beam diffracts from two gratings, creating additional beams labeled by $\textbf{k}_{2,3}$ at angle $\pm\theta$ from $\hat{\textbf{y}}$.  Each beam applies forces to atoms near the center of a linear magnetic field $\textbf{B}$.  Under certain constraints on the grating efficiency and angle, the input beam polarization, detuning, and intensity can be optimized to cool and capture atoms in two dimensions.}
\label{fig: kvectors}
\end{figure}

A circularly polarized beam with intensity $I_1$, normally incident on a grating, will diffract upwards at an angle $\theta$ from normal ($+\hat{\textbf{y}}$) with intensity $I_{\mathrm{up}}$, as shown in Fig.~\ref{fig: kvectors}.  The incident beam has $\textbf{k}_1 = -|k| \hat{\textbf{y}}$ and $s = +1$, denoting pure circular polarization.  The magnetic field $\textbf{B} = G \left( x\hat{\textbf{x}} - y\hat{\textbf{y}} \right)$ has gradient $G$ and is centered on the beam overlap region. The resulting force from beam 1 is 

\begin{equation}
\textbf{F}_1 \approx -\hbar k \frac{\Gamma}{2\pi} \frac{I_1}{I_{\mathrm{sat}}} \left[ K + C\left( -kv_y - \frac{\mu_F G}{\hbar} y \right) \right] \hat{\textbf{y}}.
\end{equation}

In general, gratings do not preserve polarization.  The diffracted beams will have a fractional intensity $P_+I_{\mathrm{up}}$ in the $s = +1$ polarization and $P_-I_{\mathrm{up}}$ in the $s = -1$ polarization.  Summing over the polarizations, the total force in $\hat{\textbf{x}}$ is

\begin{equation}
\textbf{F}_{x} \approx \hbar k C \Gamma \mbox{sin}^2\theta \frac{I_{\mathrm{up}}}{I_{\mathrm{sat}}} \bigg( k v_x + (P_--P_+)\frac{\mu_F G}{\hbar} x  \bigg)\hat{\textbf{x}}.  
\end{equation}
Similarly,

\begin{align}
\textbf{F}_{y} &\approx \hbar k \Gamma K \cos{\theta}  \frac{I_{\mathrm{up}}}{I_{\mathrm{sat}}} \nonumber \\
&+\hbar k \Gamma C \cos{\theta} \frac{I_{\mathrm{up}}}{I_{\mathrm{sat}}} \left( kv_y \cos{\theta} + 2(P_+-P_-)\frac{\mu_F G}{\hbar}y\cos{\theta} \right) \nonumber \\
&-\hbar k \frac{\Gamma}{2} \frac{I_1}{I_{\mathrm{sat}}} \left[ K + C\left( -kv_y - \frac{\mu_F G}{\hbar} y \right) \right]\hat{\textbf{y}}.
\end{align}

The constant terms (i.e. those $\propto K$) represent an intensity mismatch that will shift the trap center if not properly balanced.  In particular, a trap will only form at the magnetic field zero if

\begin{equation}
I_{\mathrm{up}} = \frac{I_1}{2\cos{\theta}}.
\label{eq: intensity}
\end{equation}
Then,
\begin{equation}
\textbf{F}_{x} \approx \hbar k C \frac{\Gamma}{2} \frac{I_1}{I_{\mathrm{sat}}} \frac{\mbox{sin}^2\theta}{\cos{\theta}} \bigg( k v_x + (P_--P_+)\frac{\mu_F G}{\hbar} x  \bigg)  \hat{\textbf{x}},
\label{eq: forcex}
\end{equation}

\begin{align}
\textbf{F}_{y}  &\approx \hbar k C \frac{\Gamma}{2} \frac{I_1}{I_{\mathrm{sat}}} \bigg( kv_y(1+\cos{\theta}) \nonumber \\
&+ \frac{\mu_F G}{\hbar} y\Big(1+(P_+-P_-)\cos{\theta}\Big) \bigg)   \hat{\textbf{y}}.
\label{eq: forcey2}
\end{align}
Note that because $\Delta$ is negative, these forces perform trapping and cooling.

Eq.~(\ref{eq: intensity}) shows the ideal intensity balance between the three beams of the 2D~GMOT.  However, a subtle distinction separates Eq.~(\ref{eq: intensity}) from the necessary grating efficiency. Gratings compress the diffracted beam area with respect to the originally incident light.  Thus, a perfectly efficient grating (i.e. 100\% of input power directed into the first order) would produce $I_{\mathrm{up}} = I_1/\cos{\theta}$.  As a result, satisfying Eq.~(\ref{eq: intensity}) requires a grating efficiency of $50\%$, independent of $\theta$.  If not, the resulting intensity imbalance manifests as an offset in the trap location from the field zero along the axis normal to the gratings \cite{McGilligan2015}.  In general, for a GMOT with $N$ diffracted beams, the ideal grating efficiency is $1/N$.  

The relatively high ($1/N = 50\%$) efficiency requirements of the 2D~GMOT preclude many grating types.  Any grating without a preferred direction would have to diffract practically all power into the $\pm 1$ orders.  Asymmetric (e.g. blazed) gratings are therefore preferable.  

Custom non-directional etched gratings have been fabricated to this standard for the 3D~GMOT \cite{Nshii, McGilligan2016, Cotter2016}, albeit with considerable design time and fabrication cost.  Such gratings often require e-beam lithography for small ($\approx 500$~nm) feature sizes.  Manufacturing large area gratings requires significant time in high-demand clean room facilities, motivating our experiment to investigate the option of using replicated blazed gratings.  

Replicated gratings are inexpensive and readily available, but confined to existing master gratings.  Additionally, replicated gratings are not designed to minimize residual specular reflections, which can undermine trap performance by driving anti-trapping transitions in the atoms.  To avoid reflected light, GMOT systems with blazed gratings have gaps between the gratings which are aligned with the central axis of the input laser.  

\begin{figure}[htbp]
  
  \begin{tabular}{@{}p{0.89\linewidth}@{\quad}p{0.\linewidth}@{}}
    \subfigimgtwo[width=\linewidth]{(a)}{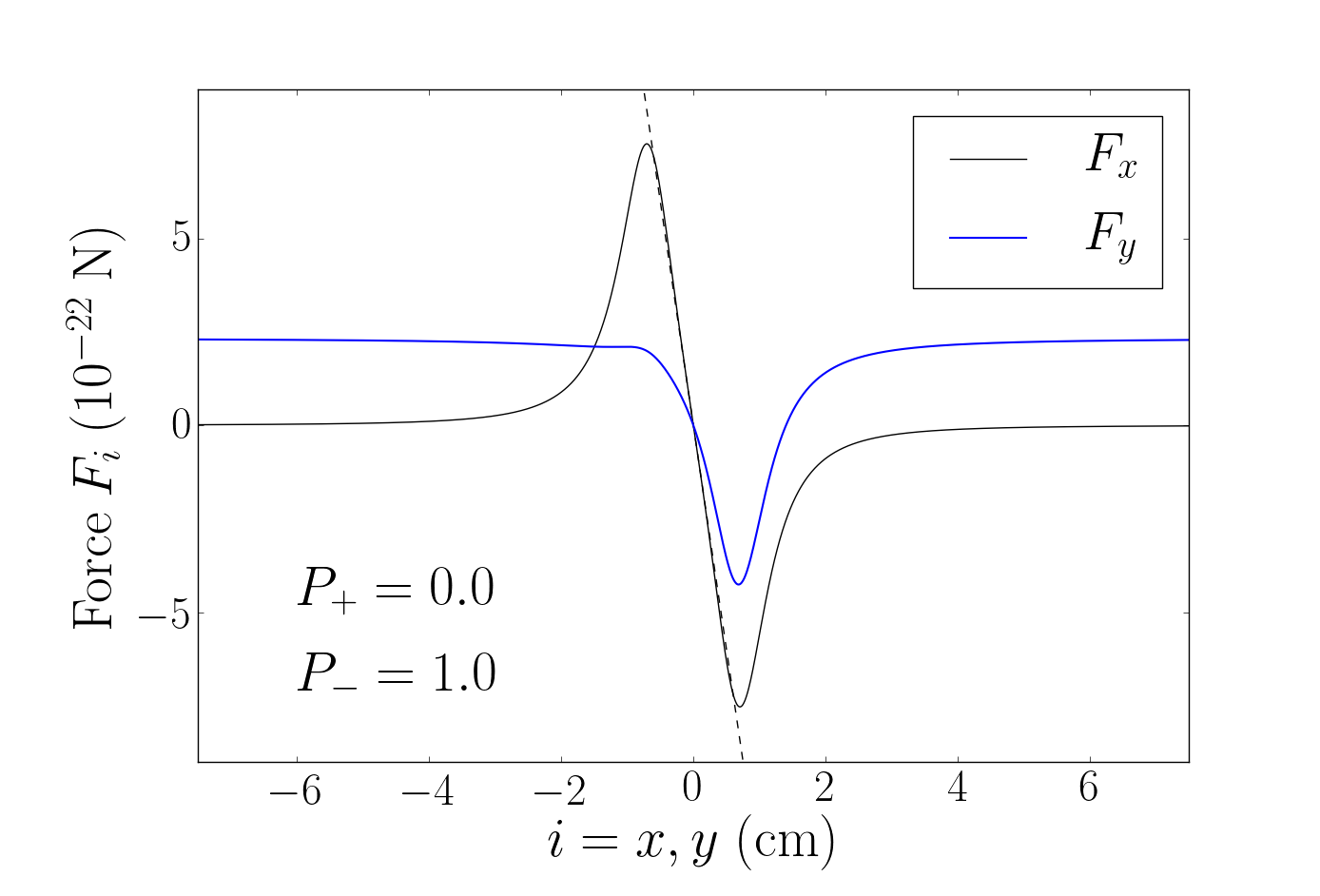} \\
    \subfigimgtwo[width=\linewidth]{(b)}{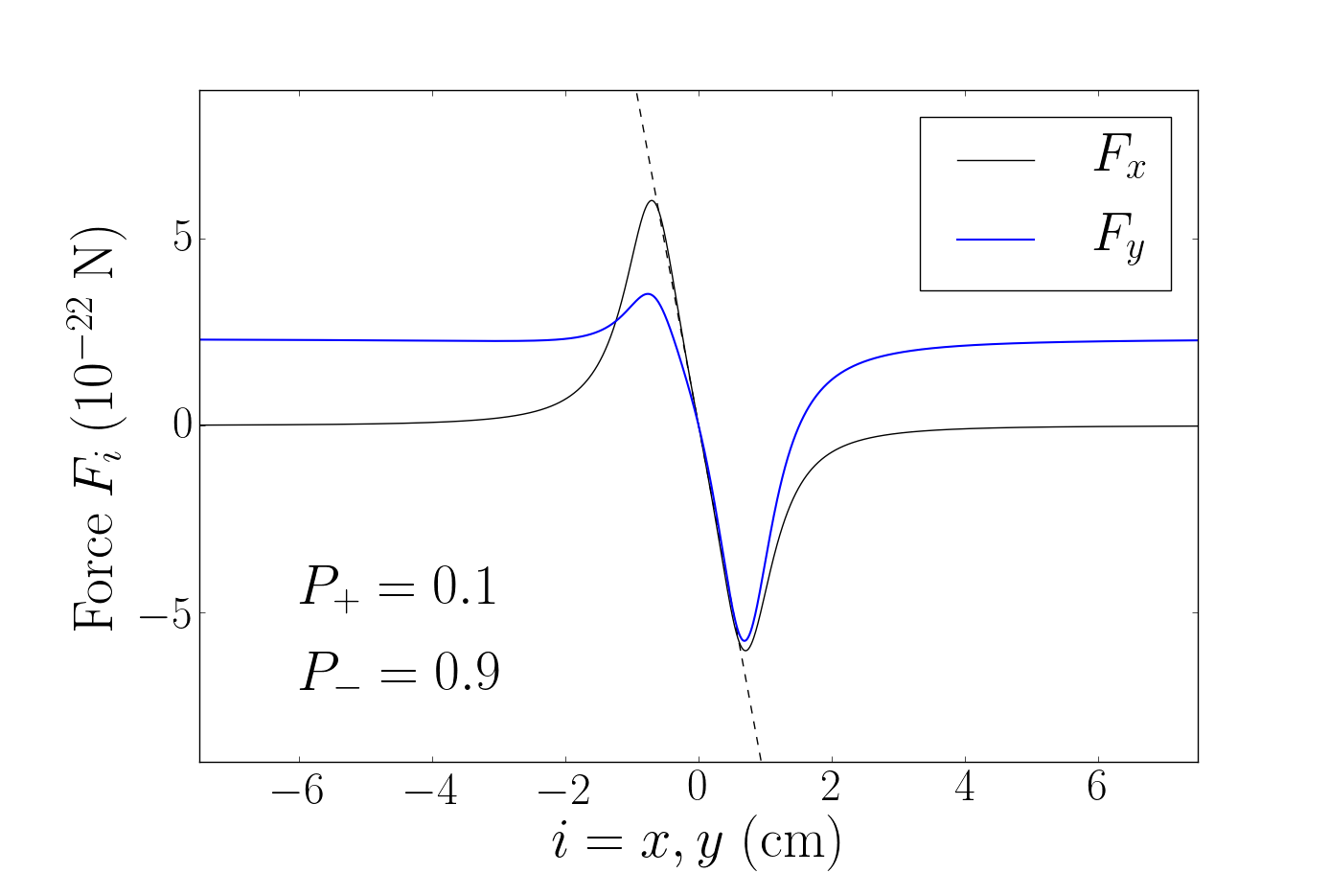} \\
    \subfigimgtwo[width=\linewidth]{(c)}{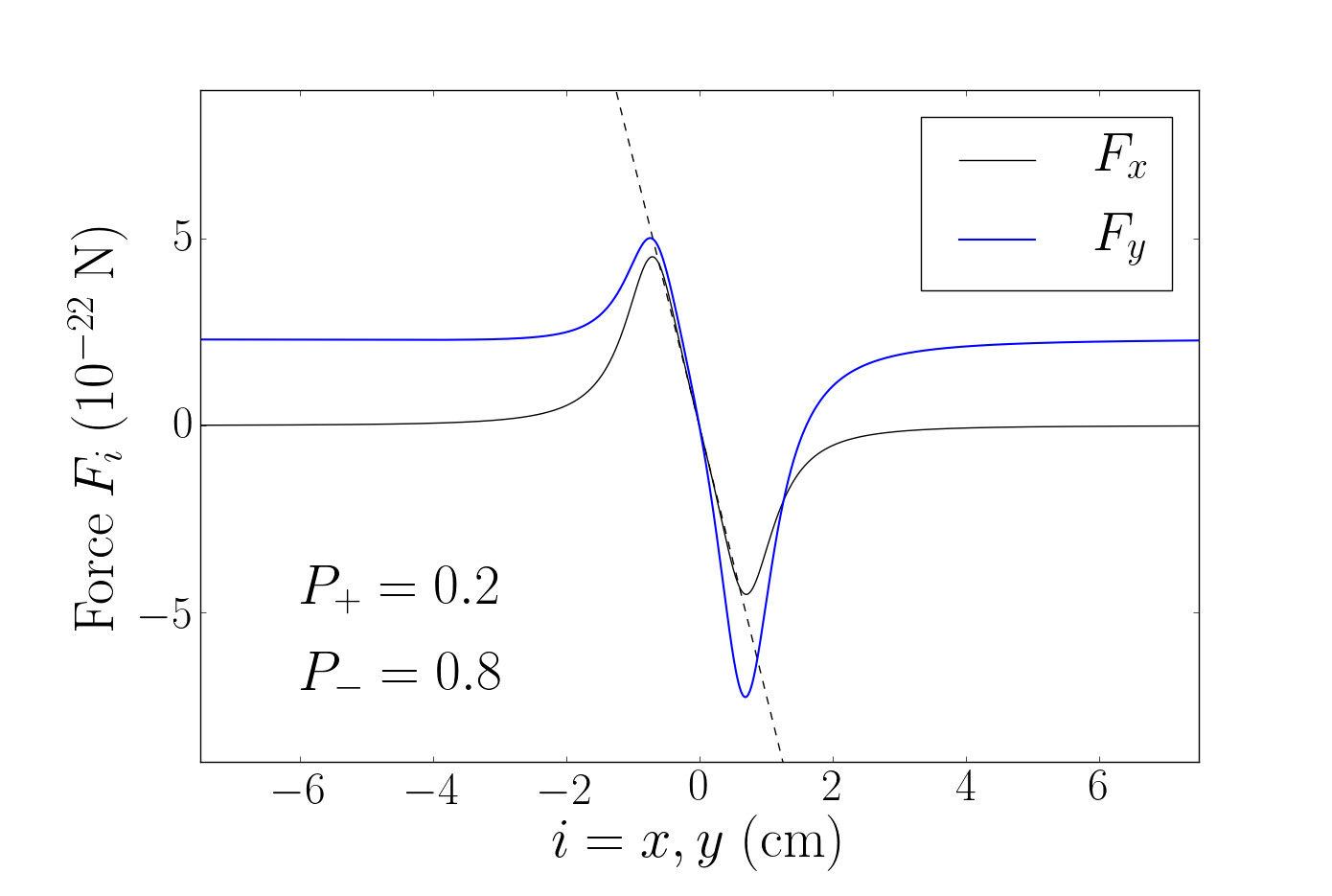} \\
    \subfigimgtwo[width=\linewidth]{(d)}{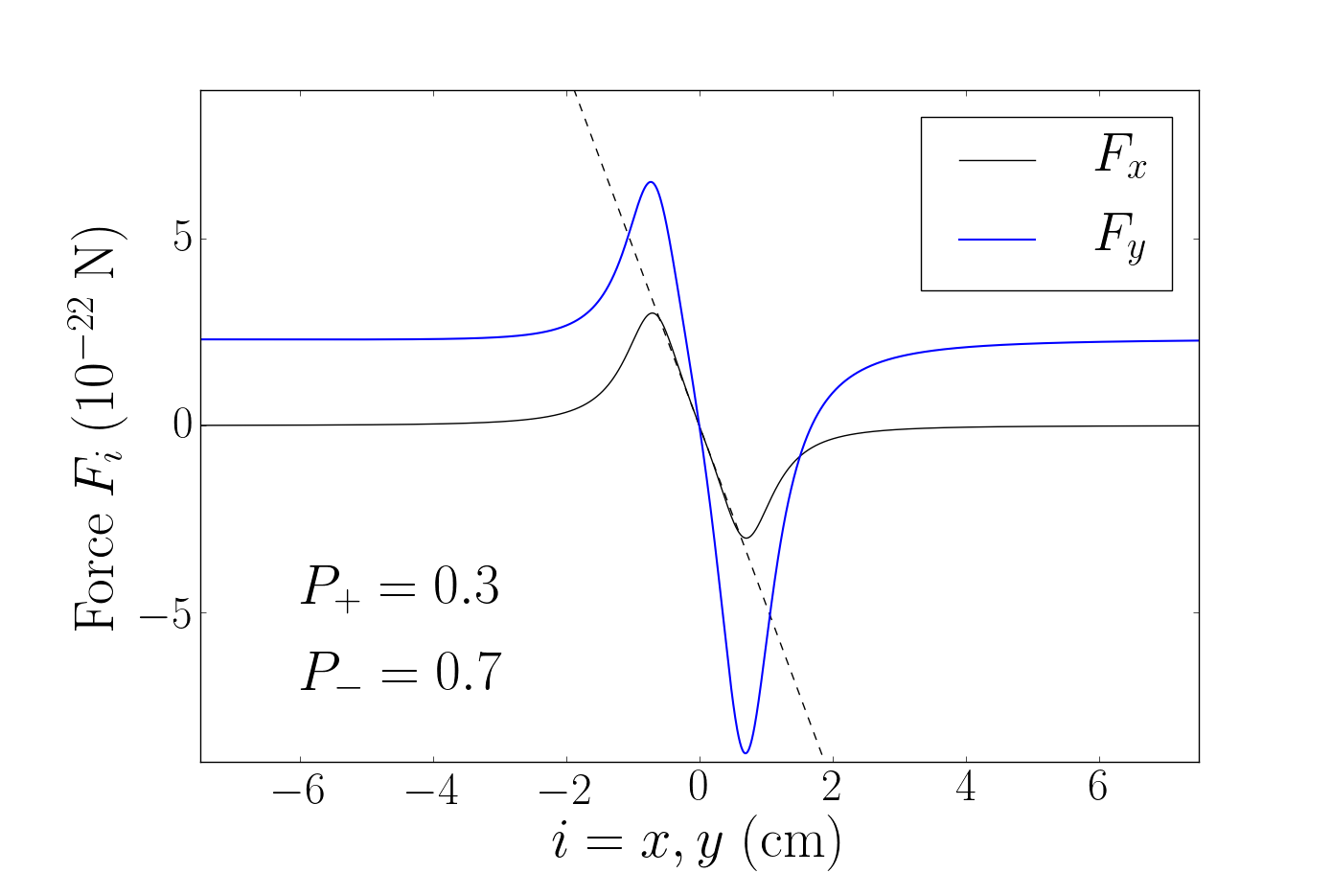} 
  \end{tabular}
  \caption{(Color online) Trapping forces in a 2D~GMOT for varying polarizations of the diffracted beams, assuming 50\% total efficiency.  Thin black curves show $\textbf{F}_x$ and thick blue curves show $\textbf{F}_y$.  Dashed black lines are the linear approximation of $\textbf{F}_x$ from Eq.~(\ref{eq: forcex}).  Plots (a)-(d) show $(P_+,P_-) = (0,1),(0.1,0.9),(0.2,0.8),$ and $(0.3,0.7)$, respectively.}
  \label{fig: forcecurves2}
\end{figure}

In addition to intensity balance, the polarization of the diffracted beams significantly effects the GMOT forces.  In particular, maximizing trapping in the $x$ direction requires $P_- = 1$ and $P_+ = 0$, as shown in Fig.~\ref{fig: forcecurves2}(a).  However, this polarization minimizes trapping in the $y$ direction.  

Fig.~\ref{fig: forcecurves2} shows the effect of imperfect polarization on the trapping forces by adjusting the ratio of $P_+$ to $P_-$ within the $50\%$ diffraction efficiency constraint.  Fig.~ \ref{fig: forcecurves2}(a)-(d) show $(P_+,P_-) = (0,1),(0.1,0.9),(0.2,0.8),$ and $(0.3,0.7)$, respectively. The linear approximation of $\textbf{F}_x$ from Eq.~(\ref{eq: forcex}) is shown as a dashed line.  The force along $y$ increases at the expense of the $x$ trapping strength.  Equal trapping strength along each axis can be achieved for $P_--P_+ = \cos{\theta}$.  For the case of $\theta = 45^\circ$, equal trapping is achieved for $P_- \approx 0.85$ and $P_+ \approx 0.15$.  

\section{Experimental Setup}

\begin{figure}[htbp]
\includegraphics[width=\linewidth]{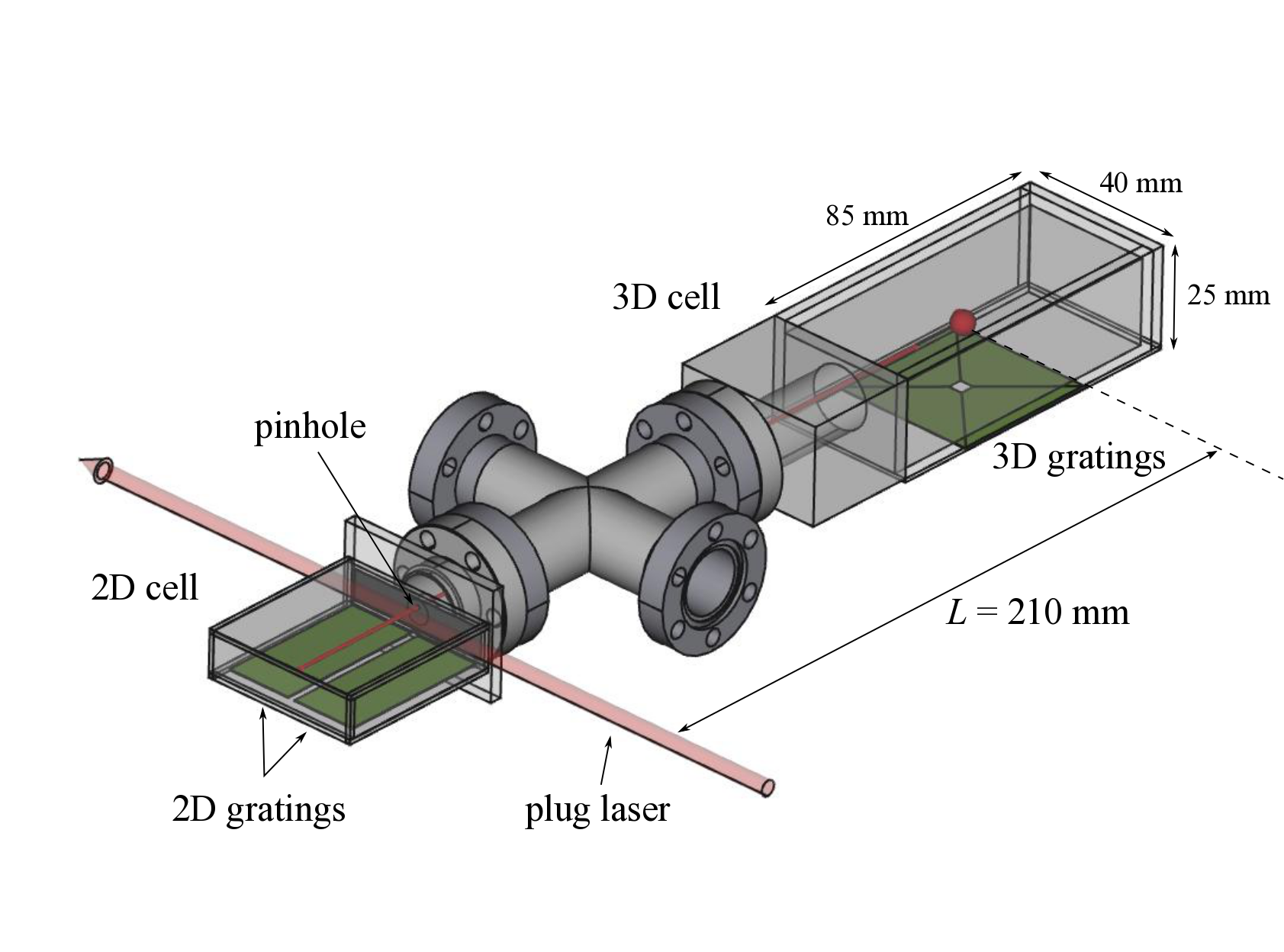}
\caption{(Color online) The experimental setup for a 2D~GMOT loading a 3D~GMOT.  Input lasers and magnetic field sources omitted for clarity.  }
\label{fig: experiment}
\end{figure}

Guided by the results of the previous section, an experiment is built to demonstrate the 2D~GMOT.  The experiment uses two epoxied glass vacuum cells \cite{Squires2016} separated by a mini-conflat flange cross, as shown in Fig.~\ref{fig: experiment}.  All cell walls are anti-reflection coated on both sides of the glass for 780 nm.  The 2D~GMOT is produced in a chamber $30\times 40\times 72$~mm$^3$, which is capped by a silicon reflector with a $1$ mm diameter pinhole.  The atom beam travels through the pinhole, then through a second filtering ($3$~mm diameter) pinhole in the copper gasket of the conflat cross.  The atoms are then collected on the opposing side of the cross by a 3D~GMOT in a $25\times 40\times 85$ mm$^3$ chamber.  

Four permanent neodymium magnets (not shown) are arranged along the corners of the 2D~GMOT chamber, creating an extended quadrupole magnetic field with a 20 G/cm gradient.  They are positioned via a three axis translation stage and a tip-tilt mirror mount to aid alignment of the 2D~GMOT with the silicon pinhole.  The 3D~GMOT magnetic fields are produced by an anti-Helmholtz coil pair, centered by cage rods that align the 3D~GMOT optics.  At 1.2 A current, they provide an axial gradient of $10$ G/cm.  

Gratings are placed outside of each vacuum chamber.  For the 2D~GMOT, two $17.5\times 38$ mm$^2$ rectangular gratings are placed with their blazes facing towards the central axis, separated by a 5 mm gap.  For the 3D~GMOT, four trapezoidal gratings are combined to produce a $38\times38$ mm$^2$ square with a $4\times 4$ mm$^2$ gap at its center.  

\begin{figure}[htbp]
\includegraphics[width=\linewidth]{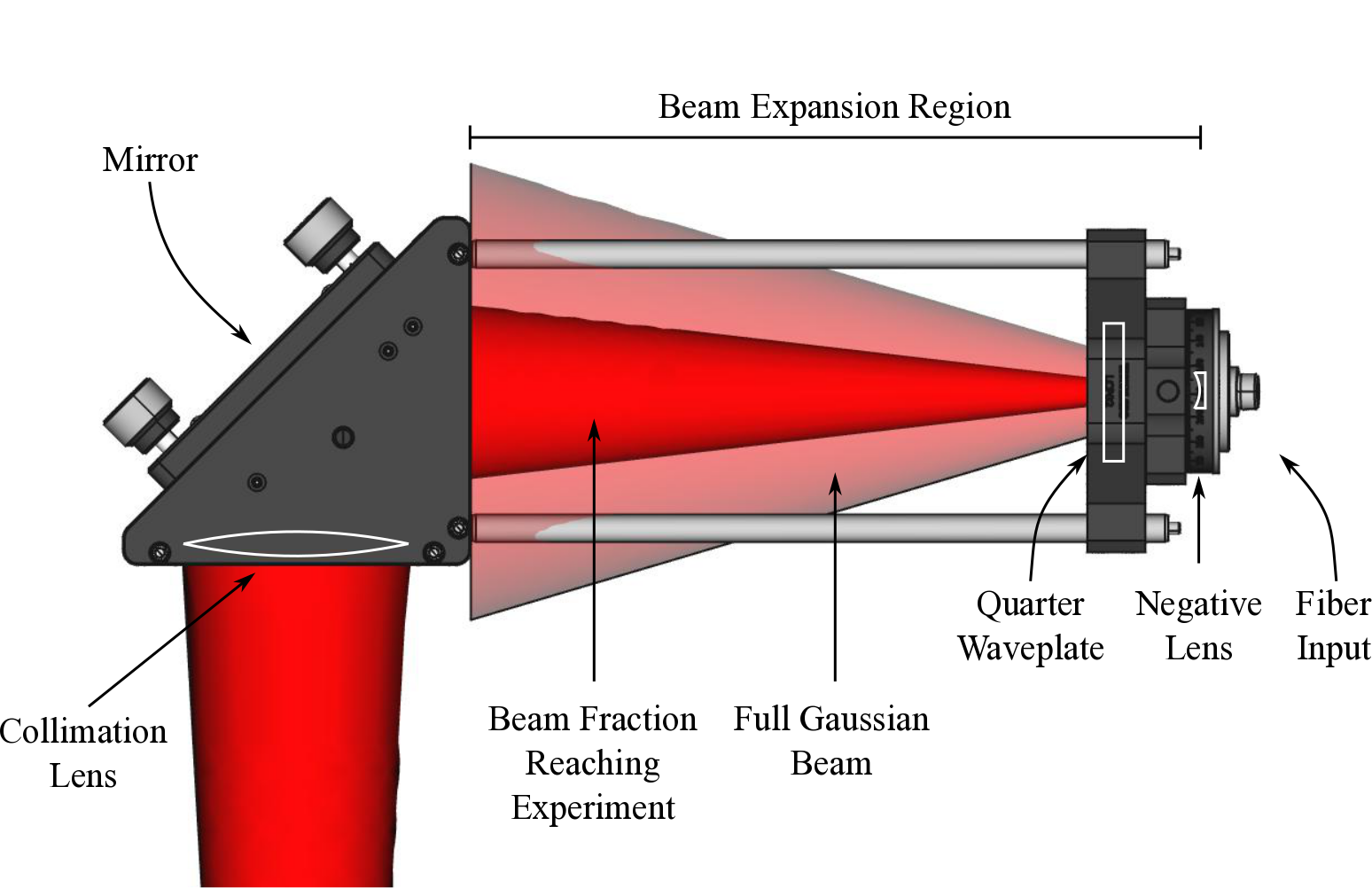}
\caption{(Color online) The optical path of the cooling light used for the 2D and 3D~GMOT.  A Gaussian beam emits from a polarization-maintaining, single mode fiber and expands through a negative lens.  A quarter waveplate provides circular polarization.  After expansion, the central, mostly uniform portion of the beam reflects from a mirror.  A final lens adjusts the remaining light's collimation. }
\label{fig: opticssetup}
\end{figure}

A single laser beam is input into each vacuum cell with $51.5$ mW red-detuned from the cooling transition for $^{87}$Rb and $18.0$ mW at the repump transition.  As shown in Fig.~\ref{fig: opticssetup}, the light is emitted from a single mode, polarization-maintaining fiber (NA = 0.12) and expanded through a negative lens ($f = -9$~mm).  A wide-angle quarter wave plate provides circular polarization to the expanding beam.  Only the central fraction of the beam is reflected towards the GMOT chamber by a two-inch mirror.  The central region has a broadly uniform intensity profile.  The reflected light passes through a two-inch lens with a 100 mm focal length.  Varying the distance from the fiber output to the final lens adjusts the collimation of the downward beam.  

The gratings are chosen using the theory presented above.  A more complete model would modify Eq.~(\ref{eq: fullforcenoapprox})  to account for the many $m_F$ states and $g_F$ factors of Rb.  These changes affect the strength of the trapping forces.  However, the derived conditions pertaining to intensity balance and polarization remain valid.  For the 2D~GMOT in particular, ideal gratings diffract at $\theta = 45^\circ$ to maximize the beam overlap area, corresponding to $\sim906$ grooves per mm.  Additionally, ideal gratings diffract circularly polarized incident light at $50\%$ efficiency while modifying the output beam to be $\approx 85\%$ circularly polarized with the opposite handedness.   

\begin{figure}[htbp]
\includegraphics[width=\linewidth]{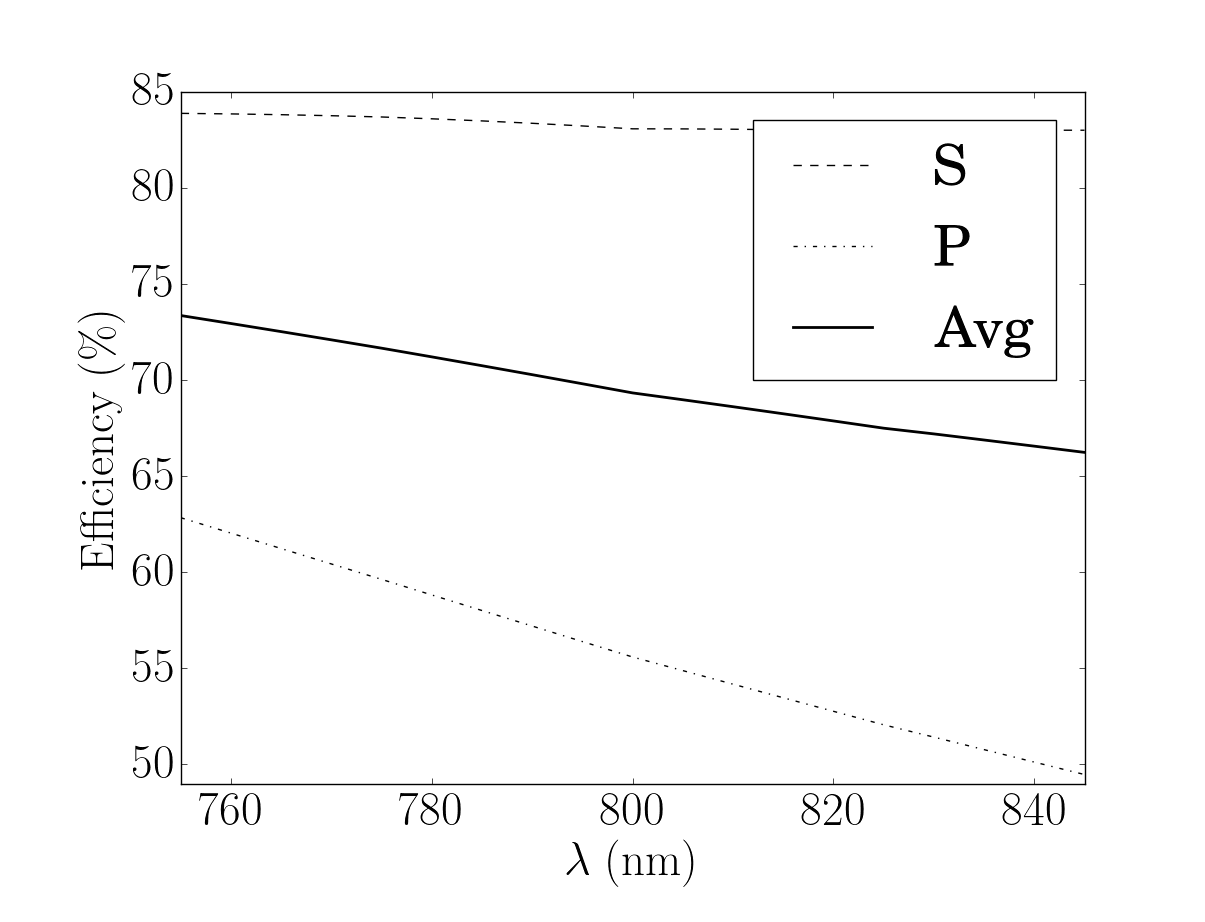}
\caption{Polarization-dependent grating efficiency as a function of wavelength at normal incidence for a grating with 830 g/mm and an 800 nm blaze.}
\label{fig: efficiencies}
\end{figure}

A commercially produced grating with 830 grooves per mm and an 800 nm blaze wavelength approximates these conditions, diffracting at $\theta = 40.3^\circ$.  Assuming light is input normal to the grating, Fig.~\ref{fig: efficiencies} shows the theoretical diffraction efficiency for incident polarization parallel and perpendicular to the groove direction.  These combine to give the average efficiency, shown as the thick solid curve.  

The circularly polarized incident beam has equal intensities of $S$ and $P$-polarized light.  Because each component diffracts differently, the output beam is elliptically polarized.   Using a Thorlabs TXP polarimeter \cite{NonEndorse}, we measure the overall diffraction efficiency at 68\% with $P_+ = 0.061$ and $P_- = 0.939$.  Because the gratings are located outside of the vacuum cell, the optical surfaces of the glass chamber modify the intensity and polarization of the diffracted beams before they reach the atoms.  As a result, the overall efficiency drops to $64\%$, with $P_+ = 0.066$ and $P_- = 0.934$.  

The non-ideal diffraction causes an intensity imbalance which can be compensated by adjusting the collimation of the input beam.  For the measurements to follow, the beam is made to focus 40 cm after the final lens, with the gratings positioned 5~cm from the lens.  Thus, in the GMOT chambers, the incident beam has an approximately uniform intensity profile with $11.0$ mW/cm$^2$ at the cooling transition and $3.8$ mW/cm$^2$ at the repump transition.  

A ``push'' beam is directed along the 2D~GMOT axis to provide enhanced longitudinal cooling, using $3.3$ mW of cooling light in a beam with a $4$ mm waist.  The beam is retro-reflected from the silicon reflector.  We refer to the 2D~GMOT with a push beam as a 2D$^+$~GMOT.  

The same gratings are used for the 3D~GMOT.  However, because the trap uses four diffracted beams, the ideal diffraction efficiency should be $1/N = 25\%$.  Accordingly, a 0.1 ND filter is placed between the 3D gratings and the chamber wall.

\section{Diagnostics}

The 3D~GMOT fluorescence is monitored using a photodiode (Thorlabs PDA100A \cite{NonEndorse}).  Light from the GMOT is collected using a $f = 25.4$ mm lens positioned $2f$ from the trap and the sensor surface.  Switching the 3D~GMOT's magnetic field on produces a rising fluorescence signal proportional to the number of captured atoms.  The 3D~GMOT atom number $N(t)$ is approximately described by the capture rate $R_{\mathrm{capture}}$ and trap lifetime $\tau_{\mathrm{trap}}$

\begin{equation}
N(t) = \tau_{\mathrm{trap}}R_{\mathrm{capture}}\left( 1-e^{-t/\tau_{\mathrm{trap}}}\right).
\label{eq: rise}
\end{equation}

\begin{figure}[htbp]
\includegraphics[width=\linewidth]{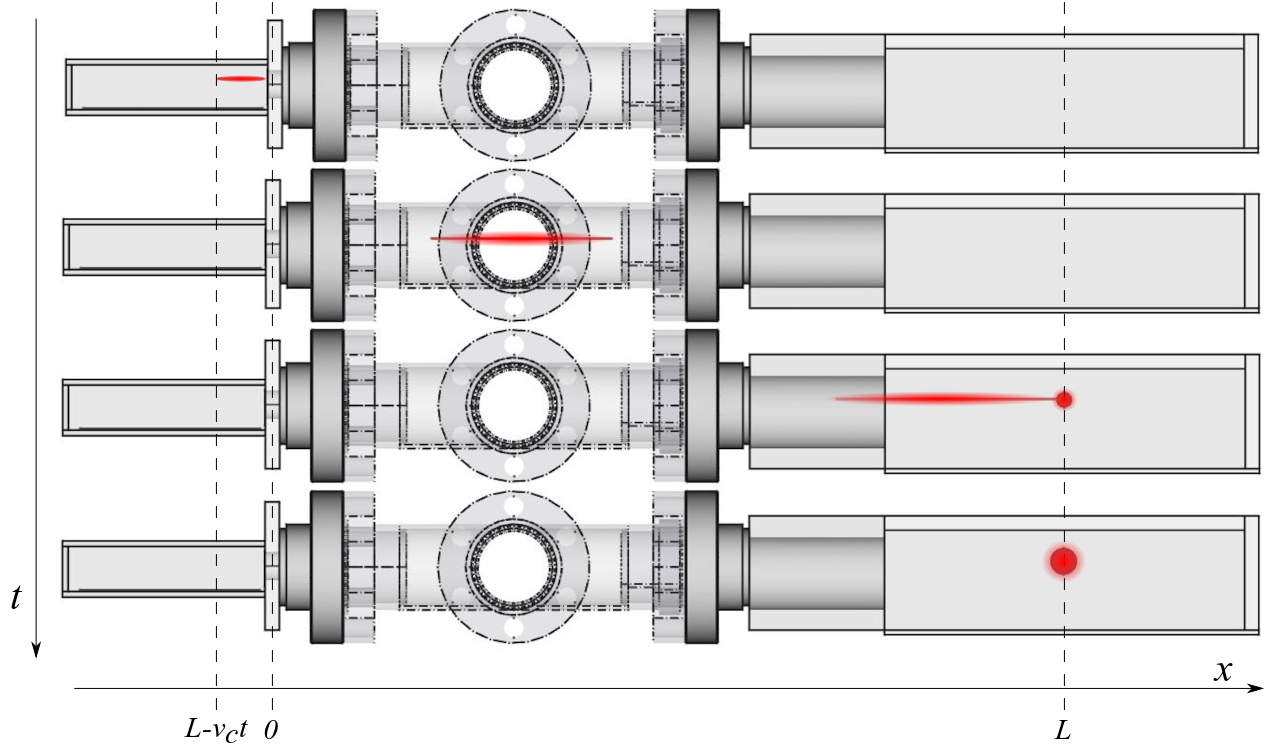}
\caption{(Color online) A short pulse of the 2D$^+$~GMOT is released at $t = 0$, traverses a distance $L$, and is captured in a 3D~GMOT, which grows as a function of time.  }
\label{fig: loadingprocessinpaper}
\end{figure}

An 8 mW ``plug'' laser beam is then positioned just before the exit pinhole, as seen in Fig.~\ref{fig: experiment}.  The plug laser acts to misalign the atomic beam from the 3D~GMOT, effectively reducing $R_{\mathrm{capture}}$ by an amount $\mathcal{R}$.  If the plug beam is turned off for a short period, the 3D~GMOT will grow as atoms traverse the distance $L$ from the exit pinhole to the capture volume of the 3D trap, as shown in Fig.~\ref{fig: loadingprocessinpaper}.  This growth is used to characterize the 2D$^+$~GMOT beam.

Analytic models for the flux of typical 2D$^+$~MOTs have been presented previously \cite{Walraven1998, Schoser2002}.  We use a simplified, closed form solution to fit the data.  Specifically, we assume the steady-state 2D$^+$~GMOT can be described as a distribution of atoms in position and velocity

\begin{equation}
\eta(z,v) = \frac{A}{\sigma \sqrt{2\pi}} \, \mbox{exp}\left( -\frac{(v-v_0)^2}{2\sigma^2} \right),
\label{eq: gaussianeta}
\end{equation}
where $A$ represents the number of atoms/m in the beam, weighted by a Gaussian distribution in velocity with peak $v_0$ and spread $\sigma$.  Thus, the density of atoms with velocities between $v_1$ and $v_2$ is $\int_{v_1}^{v_2} \eta(z,v) dv$.  

In the case of an atomic beam with a uniform speed $v_0$ (i.e.~$\sigma = 0$), no atoms reach the 3D~GMOT until $t=L/v_0$.  For $t\geq L/v_0$, a constant flux reaches the capture volume.  While $t<<\tau_\mathrm{trap}$, loss terms can be neglected and the resulting 3D~GMOT growth is linear 

\begin{equation}
N(t) =
\begin{cases}
0 & t < L/v_0 \\
\mathcal{R} \left(t-\frac{L}{v_0}\right) & L/v_0\leq t << \tau_\mathrm{trap}.
\end{cases}
\label{eq: linearmodel}
\end{equation}

A more realistic atom beam (i.e.~$\sigma > 0$) will not have such an abrupt change in $N(t)$. There is still no growth for $t<L/v_\mathrm{c}$, where $v_\mathrm{c}$ is the capture velocity of the 3D~GMOT.  But for $t\geq L/v_\mathrm{c}$, atoms with velocities between $v_\mathrm{c}$ and $v = L/t$ contribute to the 3D~GMOT number.  The velocity spread of the atom beam causes a gradual transition to linear growth given by

\begin{equation}
N(t) =
\begin{cases}
0 & t < L/v_c \\
\int_{L-v_{\mathrm{c}}t}^{0}\int_{(L-z)/t}^{v_{\mathrm{c}}} \eta(z,v) \mathop{dv}\mathop{dz} & L/v_c\leq t << \tau_\mathrm{trap}.
\end{cases}
\label{eq: model}
\end{equation}
The solution to these integrals is presented in Appendix~\ref{app: dist}.  Additionally, we show that the flux of atoms exiting the pinhole with velocities in the range $v$ to $v+dv$ is
\begin{equation}
\Phi(v)dv = \frac{A}{\sigma \sqrt{2\pi}} \,v\, \mbox{exp}\left( -\frac{(v-v_0)^2}{2\sigma^2} \right)dv.
\label{eq: flux}
\end{equation}
The total flux defines the linear slope of $N(t)$ as 

\begin{equation}
\mathcal{R} = \int_{-\infty}^{\infty} \Phi(v) \mathop{dv} =Av_0.
\end{equation}

\section{Results}

A discussion of the data processing and error analysis for the following results is provided in Appendix~\ref{app: error}.  Fig.~\ref{fig: rise} shows the rise in atom number when the 3D magnetic coils are switched on.  The solid curve is a fit to Eq.~(\ref{eq: rise}) in which $R_{\mathrm{capture}} = 1.12(3) \times 10^8$ atoms/s and $\tau_{\mathrm{trap}} = 2.20(3)$ seconds, corresponding to an upper limit on the pressure in the 3D chamber of $\approx~1\times10^{-8}$~Torr~\cite{Sackett2012}.  The steady state MOT number is $2.46(7)\times 10^8$ atoms.  

\begin{figure}[htbp]
\includegraphics[width=\linewidth]{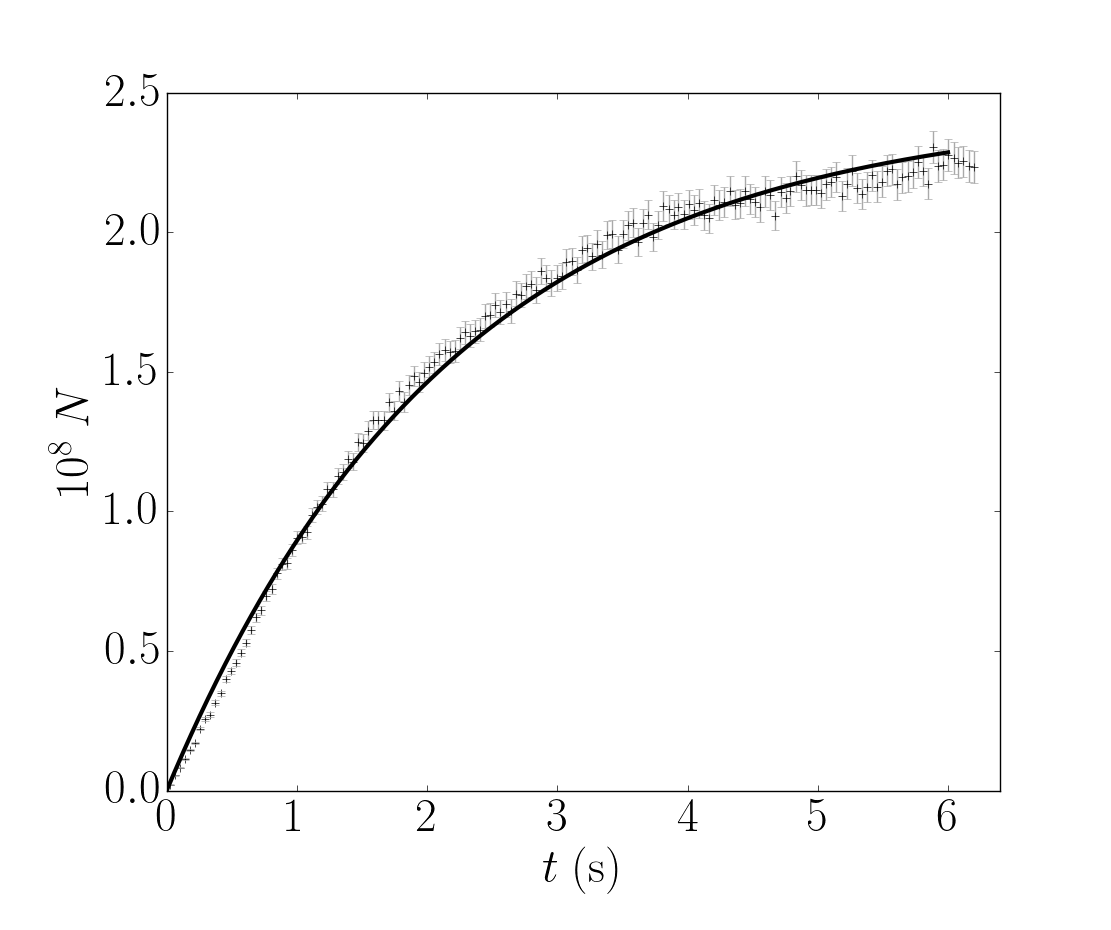}
\caption{Atom number in 3D~GMOT versus time after 3D~GMOT magnetic field is switched on.}
\label{fig: rise}
\end{figure}

The plug beam is then applied to reduce $R_{\mathrm{capture}}$.  To synchronize the subsequent time-of-flight experiment, the plug beam power is monitored with a photodiode.  The plug beam is turned off and the resulting 3D~GMOT growth recorded for total time $T_\mathrm{total}$.  Over the course of an hour, 16 independent experiments take place for each of the following: $T_\mathrm{total} =47, 97, 197, 297,$ and $397$~ms.  The longer data sets determine the linear region of $N(t)$, while the shorter data sets have greater time resolution to map the initial curvature in 3D~GMOT growth.  

\begin{figure}
  \begin{tabular}{@{}p{0.99\linewidth}@{\quad}p{0.01\linewidth}@{}}
    \subfigimg[width=\linewidth]{(a)}{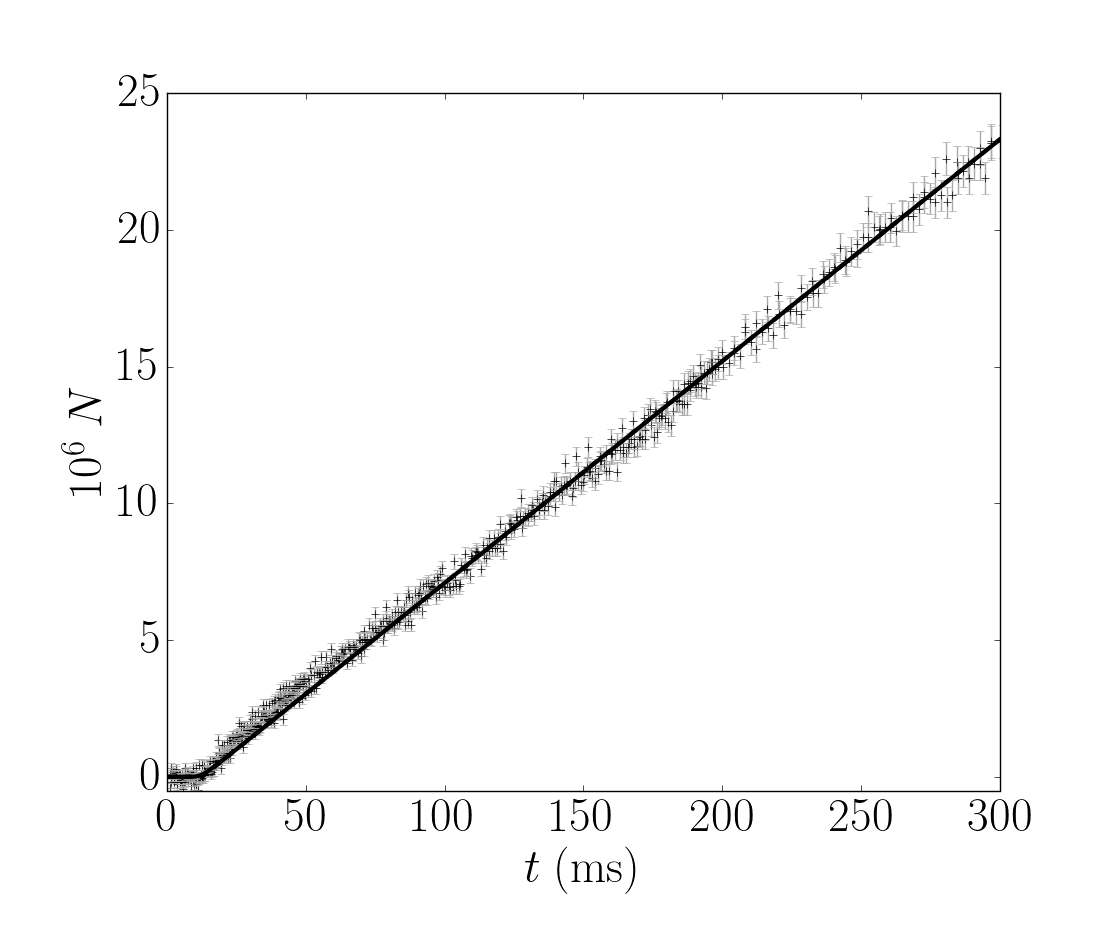} \\
    \subfigimg[width=\linewidth]{(b)}{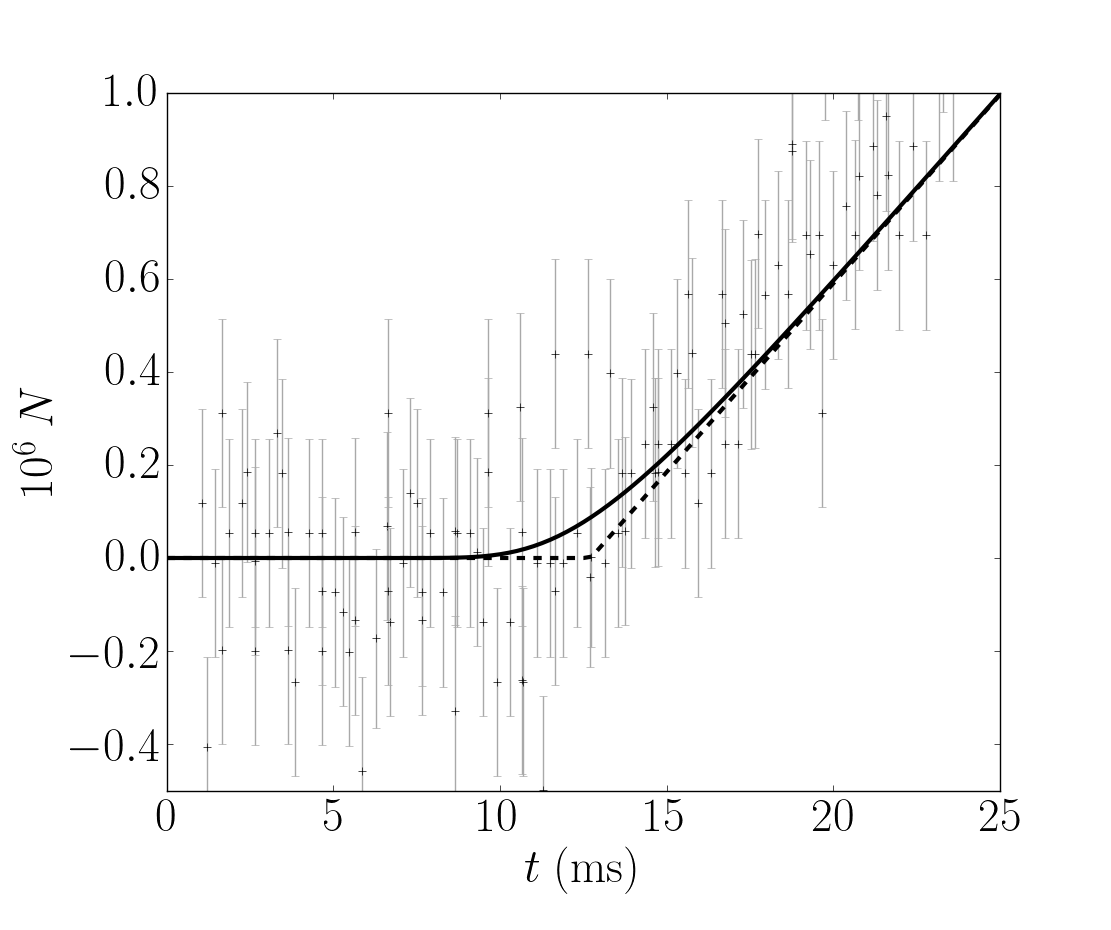} 
  \end{tabular}
\caption{Growth in 3D~GMOT atom number versus time as the plug laser beam is turned off, allowing the 2D$^+$~GMOT to load the 3D~GMOT.  The full data set is shown in (a), while the initial growth is detailed in (b).  The dashed line assumes no spread in the velocity distribution of the 2D$^+$~GMOT, as in Eq.~(\ref{eq: linearmodel}).  The solid curve is a fit using Eq.~(\ref{eq: model}).  }
\label{fig: a197ms}
\end{figure}

The combined data is shown in Fig.~\ref{fig: a197ms}.  $A$ and $v_0$ are strongly determined by the overall linearity from $t\approx100-400$~ms.  We first fit Eq.~(\ref{eq: linearmodel}) to this data, finding $A = 4.9(3)\times10^6$~atoms/m and $v_0 = 16.5(9)$~m/s.  Using these values, we fit Eq.~(\ref{eq: model}) across our entire data set to find $\sigma = 4(3)$~m/s.  The linear fit is depicted in Fig.~\ref{fig: a197ms}(b) as a dashed line, while the full fit is given as a solid curve.  

Comparing $\mathcal{R}$ to $R_{\mathrm{capture}}$, the plug beam reduces the atomic flux by $72\%$.  Additionally, it is likely that only $\approx 25\%$ of the 2D$^+$~GMOT beam actually enters the capture volume of the 3D~GMOT, assuming typical atom beam divergence as discussed in \cite{Walraven1998}.  We therefore estimate the total flux at the pinhole to be $>4\times10^8$ atoms/s.  

\section{Comparisons and Outlook}

Traditional 2D$^+$ MOT's have typical flux values near $10^9$ atoms/s \cite{Walraven1998}, and in extreme cases are as high as $10^{11}$ atoms/s \cite{Schoser2002}.  However, high flux 2D$^+$ MOT's form across 10 cm lengths or higher and saturate with laser intensities near 20 mW/cm$^2$.  By comparison, the 2D$^+$~GMOT reported here forms over a length of several mm with 11 mW/cm$^2$ laser intensity.  The short beam length is expected, as circular Gaussian beams cause the input intensity profile to vary significantly, limiting the range over which optimal cooling parameters are achieved.  Future work will employ beam shaping techniques to create a top hat intensity profile within the trap region.  A top hat intensity profile will also help make more effective use of available laser power.  

Increasing the 2D$^+$~GMOT length allows atoms with higher longitudinal velocities to be collimated into the MOT beam, increasing flux at the cost of a higher mean speed.  Additionally, length improves total output by integrating a longer capture volume.  Assuming the pressure is low enough that collisions are negligible, traditional 2D~MOT flux scales linearly with increased length \cite{Schoser2002, Serrano2006, Kellogg}.  While this experiment is not conducive to independently varying length, we expect the 2D$^+$~GMOT to scale similarly.  

Additionally, both the 2D$^+$~GMOT and 3D~GMOT should benefit from higher laser intensity, which acts to raise the capture velocity.  Prior work has shown that 2D$^+$~GMOT flux is maximal for laser intensities near $20$ mW/cm$^2$, while the 3D~GMOT atom number saturates near $50$ mW/cm$^2$ \cite{Nshii}.  Both are significantly higher than the 11~mW/cm$^2$ produced by our laser system.  Despite the difference, the loaded 3D~GMOT described here shows the highest atom number reported so far in a grating based system.  

Because this work shows the first 2D~GMOT, the system described above was designed to be large enough that time-of-flight diagnostics could be easily performed.  In future work, the 2D-to-3D GMOT system will be integrated into significantly smaller forms.  By placing the gratings within the vacuum cell and using atom chips to create the necessary magnetic fields, we are presently developing a compact, laser-cooled system.  Towards that goal, we are investigating various experimental parameters, including the grating choice, input beam polarization and collimation, capture volume, and vacuum quality.  These results suggest further GMOT development is warranted for use in field-deployable devices.

\section{Acknowledgements}

This work was supported by the Air Force Office of Scientific Research under project number 15RVCOR169.  We would like to thank the Department of Energy's Center for Integrated Nanotechnologies for its expertise in lithographic techniques and grating manufacture.  We appreciate the research group of Erling Riis at the University of Strathclyde for discussions related to GMOT design.

\bibliographystyle{apsrev}
\bibliography{paperref}

\begin{thebibliography}{33}
\bibitem[{Non()}]{NonEndorse}
\bibinfo{howpublished}{Reference to any specific commercial product or
  corporation name is for the information and convenience of the public, and
  does not constitute endorsement, recommendation, or favoring by the authors
  or the U.S. Government.}
\end{thebibliography}

\newpage
\appendix
\section{2D~GMOT Derivation}

The average force from the $j^{th}$ beam is 
\begin{equation}
\textbf{F}_j \approx \hbar \textbf{k}_j \frac{\Gamma}{2} \frac{I_j}{I_{\mathrm{sat}}} \left[ K + C\left( \textbf{k}_j \cdot \textbf{v} - \frac{\mu_F s}{\hbar} \frac{\textbf{k}_j \cdot \textbf{B}}{|\textbf{k}_j|} \right) \right].
\end{equation}
The magnetic field is $\textbf{B} = G\left( x\hat{\textbf{x}}-y\hat{\textbf{y}} \right)$.  The three beams have $\textbf{k}$ vectors,
$$
\textbf{k}_1 = -k\hat{\textbf{y}}
$$
with polarization $s = +1$ and
$$
\textbf{k}_2 = k\left( \sin{\theta}\hat{\textbf{x}} + \cos{\theta}\hat{\textbf{y}} \right)
$$
$$
\textbf{k}_3 =k\left( -\sin{\theta}\hat{\textbf{x}} + \cos{\theta}\hat{\textbf{y}} \right)
$$
with fraction $P_+$ in the $s=+1$ polarization and the remainder $P_-$ in the $s=-1$ polarization, where $\theta$ is the diffraction angle from the $+\hat{\textbf{y}}$ axis.  
\subsection{Beam 1}
For the input beam,

\begin{align}
\textbf{F} &\approx \hbar \textbf{k}_1 \frac{\Gamma}{2} \frac{I_1}{I_{\mathrm{sat}}} \left[ K + C\left( \textbf{k}_1 \cdot \textbf{v} - \frac{\mu_F s}{\hbar} \frac{\textbf{k}_1 \cdot \textbf{B}}{|\textbf{k}_1|} \right) \right] \nonumber\\
&\approx -\hbar k \frac{\Gamma}{2} \frac{I_1}{I_{\mathrm{sat}}} \left[ K + C\left( -kv_y - \frac{\mu_F G}{\hbar} y \right) \right] \hat{\textbf{y}}. 
\end{align}

\subsection{Beam 2}
For the $s=+1$ fraction of the second beam,

\begin{align}
\textbf{F} &\approx \hbar k\langle \sin{\theta},\cos{\theta} \rangle \frac{\Gamma}{2} \frac{P_+I_2}{I_{\mathrm{sat}}} \bigg[ K + C\bigg( k v_x \sin{\theta} \nonumber\\
&+ kv_y \cos{\theta} - \frac{\mu_F G}{\hbar} x \sin{\theta} + \frac{\mu_F G}{\hbar}y\cos{\theta} \bigg) \bigg]. 
\end{align}
For the $s=-1$ fraction of the second beam,

\begin{align}
\textbf{F} &\approx \hbar k\langle \sin{\theta},\cos{\theta} \rangle \frac{\Gamma}{2} \frac{P_-I_2}{I_{\mathrm{sat}}} \bigg[ K + C\bigg( k v_x \sin{\theta} \nonumber\\
&+ kv_y \cos{\theta} + \frac{\mu_F G}{\hbar} x \sin{\theta} - \frac{\mu_F G}{\hbar} y\cos{\theta}  \bigg) \bigg]. 
\end{align}

\subsection{Beam 3}
For the $s=+1$ fraction of the third beam,

\begin{align}
\textbf{F} &\approx \hbar k\langle -\sin{\theta},\cos{\theta} \rangle \frac{\Gamma}{2} \frac{P_+I_3}{I_{\mathrm{sat}}} \bigg[ K + C\bigg( -k v_x \sin{\theta} \nonumber\\
&+ kv_y \cos{\theta} + \frac{\mu_F G}{\hbar} x \sin{\theta} + \frac{\mu_F G}{\hbar}y\cos{\theta} \bigg) \bigg].
\end{align}
For the $s=-1$ fraction of the third beam,

\begin{align}
\textbf{F} &\approx \hbar k\langle -\sin{\theta},\cos{\theta} \rangle \frac{\Gamma}{2} \frac{P_-I_3}{I_{\mathrm{sat}}} \bigg[ K + C\bigg( -k v_x \sin{\theta} \nonumber\\
&+ kv_y \cos{\theta} - \frac{\mu_F G}{\hbar} x \sin{\theta} - \frac{\mu_F G}{\hbar} y\cos{\theta}  \bigg) \bigg].
\end{align}

\subsection{Total Forces}
Combining the contributions of each beam in the $\hat{\textbf{x}}$ direction with $I_{\mathrm{up}} = I_1 = I_2$ and $P_++P_- = 1$,

\begin{equation}
\textbf{F}_{tot,x} \approx \hbar k C \Gamma \mbox{sin}^2\theta \frac{I_{\mathrm{up}}}{I_{\mathrm{sat}}} \bigg( k v_x + (P_--P_+)\frac{\mu_F G}{\hbar} x  \bigg).  
\end{equation}

Similarly,

\begin{align}
\textbf{F}_{tot,y} &\approx \hbar k \cos{\theta} \frac{\Gamma}{2} \frac{I_{\mathrm{up}}}{I_{\mathrm{sat}}} \bigg(2K+\bigg[ 2C\bigg( kv_y \cos{\theta} \nonumber\\
&~~~~~~~~~~~~+ 2(P_+-P_-)\frac{\mu_F G}{\hbar}y\cos{\theta} \bigg) \bigg]\bigg) \nonumber\\
&-\hbar k \frac{\Gamma}{2} \frac{I_1}{I_{\mathrm{sat}}} \bigg[ K + C\bigg( -kv_y - \frac{\mu_F G}{\hbar} y \bigg) \bigg].
\end{align}
For the constant terms (i.e. those $\propto K$) to cancel, $I_{\mathrm{up}} = I_1/2\cos{\theta}$.  Then,
\begin{equation}
\textbf{F}_{tot,y} \approx \hbar k C \frac{\Gamma}{2} \frac{I_1}{I_{\mathrm{sat}}} \bigg( kv_y(1+\cos{\theta}) + \frac{\mu_F G}{\hbar} y(1+(P_+-P_-)\cos{\theta}) \bigg).
\end{equation}

\newpage
\section{GMOT Distribution Derivation}
\label{app: dist}

\subsection{Beam Distribution}

When the plug beam is pulsed off for a short period, a small packet of atoms from the 2D$^+$~GMOT is allowed to pass through the pinhole, across a distance $L$, to the 3D~GMOT trapping region.  If the atoms from the beam packet are slower than the capture velocity $v_{\mathrm{c}}$, they will be collected into the 3D~GMOT, which will grow with increased atom number.  The process is illustrated in Fig.~\ref{fig: loadingprocessinpaper}.  

Define the pinhole to be at $z=0$.  Assume that at $t=0$, the atoms are distributed uniformly behind the pinhole $(z<0)$ with no atoms past the pinhole $(z>0)$.  Assume the atoms have a Gaussian distribution in velocity.  The number of atoms between $z$ and $z+dz$ with velocities between $v$ and $v+dv$ is given by 

\begin{equation}
\eta(z,v)\mathop{dz} \mathop{dv} = \frac{A}{\sigma \sqrt{2\pi}} \mbox{exp}\left( -\frac{(v-v_0)^2}{2\sigma^2} \right)\mathop{dz}\mathop{dv},
\end{equation} 
where $v_0$ is the peak velocity of the distribution and $\sigma$ is the velocity spread.  $A$ represents the number of atoms/m, which is weighted by a normal distribution in velocity.  The total number of atoms with initial positions between $z_1$ and $z_2$ with velocities between $v_1$ and $v_2$ is 

\begin{equation}
N = \int_{z_2}^{z_1}\int_{v_1}^{v_2} \eta(z,v) \mathop{dv} \mathop{dz}.
\label{eq: basicN}
\end{equation}

The 3D~GMOT size at time $t$ is proportional to the number of atoms that reach the point $z = L$ with velocities less than $v_{\mathrm{c}}$ at or before time $t$.  In other words, an atom at position $z$ must travel at least $L+|z|$ in time $t$.  Accordingly, the minimum velocity that reaches the 3D~GMOT by time $t$ is $v_1 = (L+|z|)/t$.  The velocity range that can effect the 3D~GMOT at time $t$ is then $[v_1,v_2] = [(L-z)/t,v_{\mathrm{c}}]$.  

At $t=0$, no atoms exist past the pinhole, so $z_2 = 0$.  The fastest atom capable of being trapped is $v_{\mathrm{c}}$, and it can only travel a distance $v_{\mathrm{c}}t$ in time $t$.  The fastest atom can have an initial position no further behind the pinhole than $z_1 = L-v_{\mathrm{c}}t$.  Using these limits, the total number of atoms that reach the 3D~GMOT by time $t$ is

\begin{align}
&N(t) = \int_{L-v_{\mathrm{c}}t}^{0}\int_{(L-z)/t}^{v_{\mathrm{c}}} \eta(z,v) \mathop{dv} \mathop{dz} \nonumber\\
&= \frac{A}{\sigma \sqrt{2\pi}}\int_{L-v_{\mathrm{c}}t}^{0}\int_{(L-z)/t}^{v_{\mathrm{c}}} \mbox{exp}\left( -\frac{(v-v_0)^2}{2\sigma^2} \right) \mathop{dv} \mathop{dz}\nonumber \\
&=\frac{A}{2} \int_{L-v_{\mathrm{c}}t}^0  \mbox{erf}\left[ \frac{v_0-\frac{L-z}{t}}{\sigma\sqrt{2}} \right]dz - \frac{A}{2} \int_{L-v_{\mathrm{c}}t}^0\mbox{erf}\left[ \frac{v_0-v_{\mathrm{c}}}{\sigma\sqrt{2}} \right] dz.  \nonumber \\
&= A \frac{\sigma t}{\sqrt{2\pi}} \left( \mbox{exp}\left[ -\frac{(\frac{L}{t}-v_0)^2}{2\sigma^2} \right] - \mbox{exp}\left[ -\frac{(v_0 - v_{\mathrm{c}})^2}{2\sigma^2} \right] \right) \nonumber \\
&+\frac{A}{2} (v_0t-L)\,\mbox{erf}\left[ \frac{v_0 - \frac{L}{t}}{\sigma\sqrt{2}} \right] - \frac{A}{2}t(v_0 -v_{\mathrm{c}})\,\mbox{erf}\left[ \frac{v_0 -v_{\mathrm{c}}}{\sigma\sqrt{2}} \right] \nonumber\\
&+\frac{A}{2}(L-v_{\mathrm{c}}t) \,\mbox{erf}\left[ \frac{v_0 - v_{\mathrm{c}}}{\sigma \sqrt{2}} \right].  
\end{align}

\subsection{Flux Distribution}
Integrating $\eta(z,v)$ from $v=v'$ to $v'+dv$ gives the density of atoms in the 2D$^+$~GMOT with velocities in that infinitesimal range.  Further integrating from $z=-v'(t+dt)$ to $-v't$ gives the number of those atoms exiting the pinhole between times $t$ and $t+dt$.  

\begin{align}
dN(v',t) &= \int_{-v'(t+dt)}^{-v't}\int_{v'}^{v'+dv} \eta(z,v) \mathop{dv} \mathop{dz}\nonumber\\
&=\int_{-v'(t+dt)}^{-v't}\left[\eta(z,v')dv\right] dz \nonumber\\
&=A\int_{-v'(t+dt)}^{-v't} \left[ \frac{1}{\sigma \sqrt{2\pi}} \mbox{exp}\left( -\frac{(v'-v_0)^2}{2\sigma^2} \right)dv\right] dz\nonumber\\
&= \frac{Av'}{\sigma \sqrt{2\pi}} \mbox{exp}\left( -\frac{(v'-v_0)^2}{2\sigma^2} \right) dvdt.  
\end{align} 
Dropping the primes, the flux of atoms in a narrow range of velocities between $v$ and $v+dv$ is

\begin{equation}
\Phi(v)dv = \frac{dN}{dt} = \frac{A}{\sigma \sqrt{2\pi}} \,v\, \mbox{exp}\left( -\frac{(v-v_0)^2}{2\sigma^2} \right) dv,
\end{equation}
which peaks when $v = (v_0 \pm \sqrt{v_0^2 + 4\sigma^2})/2$.  The total flux that was blocked by the plug beam is 

\begin{equation}
\mathcal{R} = \int_{-\infty}^{\infty} \Phi(v) \mathop{dv} =Av_0.
\end{equation}

\newpage
\section{Data Processing and Error Analysis}
\label{app: error}

As the 3D~GMOT grows, its fluorescence is recorded as a series of voltage signals from the photodiode.  The $i^\mathrm{th}$ measured signal $S_i$ is converted to atom number $N_i$ with associated error $w_i$ given by the following uncertainties:

\begin{enumerate}
\item The photodiode monitoring the 3D GMOT fluorescence has $799~\mu$V root-mean-square noise at the 40 dB gain setting. 
\item The measured voltage could result from either a change in atom number or a variation in the scattering rate $R_\mathrm{sc}$, which depends on laser intensity and detuning.  On the time-scale of this experiment, $\Delta = -10.1(1)$~MHz and $I = 11.03(2)$~mW/cm$^2$.  These laser fluctuations lead to an additional $0.9\%$ uncertainty in the measured voltage.  
\item Conversion of the measured voltage to atom number is imprecise.  The atom number is given by \cite{LewandowskiThesis} as

\begin{equation}
N_i = \frac{4\pi S_i}{G \Omega \beta E_{\mathrm{photon}} R_\mathrm{sc} (T_\mathrm{glass})^m},
\end{equation}
for the detector gain $G$ and responsivity $\beta$, imaged solid angle $\Omega$, and photon energy $E_\mathrm{photon}$.  $T_\mathrm{glass}$ is the transmissivity of the $m$ optical surfaces in the imaging setup.  Accounting for the relevant uncertainties in these values, the conversion is $N_i \approx [1.61(4)\times10^9] S_i$.  
\end{enumerate}

The acquired data is binned in time every $M = 10$ points, such that the $k^\mathrm{th}$ bin is represented by mean time 

\begin{equation}
\bar{t}_k = \frac{1}{M}\sum_{l=1}^M t_{i},
\end{equation}
weighted average signal
\begin{equation}
\bar{N}_k = \frac{\sum_{l=1}^M \frac{N_i}{w_i^2}}{\sum_{l=1}^{M} \frac{1}{w_i^2}},
\end{equation}
and weighted average error
\begin{equation}
\bar{w}_k = \sqrt{\frac{1}{\sum_{l=1}^{M} \frac{1}{w_i^2}}},
\end{equation}
where $i = Mk+l$.

\end{document}